\documentstyle[sprocl,psfig]{article}

\def\ra{\rightarrow}
\def\lra{\longrightarrow}

\def\vr{{\bf r}}
\def\vv{{\bf v}}
\def\vu{{\bf u}}

\def\be{\begin{equation}}
\def\ee{\end{equation}}
\def\bea{\begin{eqnarray}}
\def\eea{\end{eqnarray}}

\begin{document}

\title{BOGOLIUBOV CHOH UHLENBECK THEORY:\\
CRADLE OF MODERN KINETIC THEORY}

\author{ MATTHIEU H. ERNST }

\address{Institute for Theoretical Physics,\\
Universiteit Utrecht, The Netherlands\\
email: m.h.ernst@fys.ruu.nl\\
\vspace{5mm}
(dedicated to the memory of Soon-Tahk Choh)}

\maketitle
\abstracts{
The Choh Uhlenbeck equation was the first successful generalization of the Boltzmann equation to include triple collisions in a systematic manner.
 Detailed analysis of its triple collision term has led to the important concept of {\it ring collisions}, which create dynamic correlations
between particles with a common collision history. 
This review shows how the ring collisions have been at the root of most major developments in kinetic theory of the last 40 years, both for continuous 
fluids as well as for lattice gas automata. In present times they have 
become the standard tool to analyze long spatial and temporal 
correlations in simple and complex fluids.}

\section{Historical perspective}
In the sixties many efforts were made to generalize the Boltzmann 
equation to
dense liquids. The historical importance
of the Bogoliubov Choh Uhlenbeck theory and the Choh Uhlenbeck
equation is well expressed by the following
quote from Brush' book~\cite{Brush} on the history of kinetic theory: ``At 
Michigan,
Soon-Tahk Choh, a Korean student of Uhlenbeck, was doing apparently 
similar
but ultimately more fruitful work. Early in 1958 he completed his
dissertation, {\it The kinetic theory of phenomena in dense gases}~\cite{Choh},
a  work never
published but frequently cited  by later physicists who heard about 
it''.

In an effort to put Choh's work in perspective I briefly 
review 
some important events, relevant for the development of the kinetic 
theory
of dilute and dense fluids. In a small variation on Cohen's 
classification in his {\it Fifty years of kinetic theory} \cite{Coh-Berl} 
I divide the history of kinetic theory into four periods:
\begin{tabbing}
$xxx$\=xxxxxxxxxxxxxxxxxxxxx\=xxxxxxxxxxxxxxxx\kill
$\bullet$ \> Classic Era \> 1855-1945\\
$\bullet$ \> Renaissance \> 1946-1964\\
$\bullet$ \> Modern Era  \> 1965-1985\\
$\bullet$ \> Post-modern Era \> $>$ 1985
\end{tabbing}

The selection of important developments is based on 
Brush's book, the overview in Choh's thesis, and on review papers of personal preference \cite{DvB,EHD,Pom+R,Coh-Berl}.

The {\it Classical Era} started with Maxwell's derivation in 1859 of the 
velocity distribution in thermal equilibrium: the 
birth of kinetic theory. Soon afterwards, in 1872 Boltzmann formulated his famous equation for the
time evolution of the single particle distribution function, $f(x,t) 
=
f(\vr , \vv,t)$, in a {\it dilute gas} outside 
equilibrium,
\be
\partial_t f +  \vv \cdot \mbox{\boldmath$ \nabla$} f = I(ff) .
\ee
The collision term on the right hand side, which is quadatic in $f$, 
represents
the change in $f$ through binary collisions. It is based on the 
assumption of
{\it molecular chaos}, expressing the absence of precollision 
correlations between
two particles involved in a binary collision. 
In other words, the Boltzmann equation takes only
 sequences of uncorrelated binary collisions 
into account. This implies that the pair 
distribution function $f^{(2)}(x_1 x_2,t)$, just before particles 1 
and 2 enter each others interaction range, factorizes into
\be
f^{(2)}(x_1 x_2,t) \ra f(x_1,t) f(x_2,t) \simeq 
f( \vr , \vv ,t) f( \vr , \vv_2,t).
\ee
The Boltzmann equation shows the
arrow of time; it is an irreversible equation, which is lacking the 
time
reversal invariance of the mechanical equations of motion. In fact,
$f(x,t)dx$ represents the {\it average} number of particles in a 
small
element, $dx = d\vr d\vv$ of phase space. 
The Boltzmann equation {\it neglects fluctuations} and can be 
considered
as a mean field theory. Boltzmann was able to prove the $H$--theorem and 
discuss
the approach of $f(x,t)$ towards thermal equilibrium, described by
Maxwell's velocity distribution function.

The remaining problems were to derive the Navier Stokes equation and 
to calculate
the transport coefficients. The possibility of an explicit 
calculation of
viscosity and heat conductivity was realized in 1917 when 
Chapman and Enskog developed the {\it Chapman Enskog method} for solving the 
Boltzmann
equation. This multi-time scale method assumes that transport 
processes, which
occur on time scales large compared to the mean free time $t_0$, are 
described
by the {\it normal solution}, $f(\vr , \vv ,t) = f(\vv  | n, \vu , 
T)$, which 
depends on space and time variables $(\vr ,t)$ only through the 
first five
moments of the distribution function, $a(\vr ,t) = 
\{ n(\vr ,t), \vu (\vr ,t),
T(\vr ,t)\}$, i.e. the local density $n$, the local flow
velocity $\vu $ and the local temperature $T$.

In this time regime, the temporal and spatial variations over a mean 
free 
time $t_0$ and a mean free path $\ell_0 = \overline{v}t_0$, provide a 
small parameter, the {\it non-uniformity parameter}, $\mu \sim 
\ell_0 \nabla$, which can be used as a systematic 
expansion parameter~\cite{Coh-Berl}. 
Time derivatives, $\partial_t f \sim (\partial f /\partial 
a) \partial_t a$, can be
calculated by carrying out a $\mu$-expansion of the local 
conservation laws for
$a(r,t)$, where $\partial_t a$ is at least of {\cal O}($\mu$). Then, 
the Boltzmann equation can be solved by a $\mu$-expansion, {\it i.e.}
 $f= f_0 + 
\mu f_1 + \dots$, where $f_0 = n(m/2\pi kT)^{d/2} \exp 
[ -m| \vv - \vu |^2 /  2kT]$  is the local 
equilibrium distribution with $d$ the dimensionality of the system.
The Chapman Enskog method yields a linear integral equation for 
the
deviation $f_1$, which is proportional to the gradients of $T$ and 
$\vu $, and provides explicit expressions for the heat
conductivity and viscosity.

In 1922 Enskog proposed  a rather
accurate~\footnote{`rather accurate', that is in moderately dense 
systems where the mean free path $\ell_0$ is larger than the hard 
core diameter $\sigma$. In dense liquids, where $\ell_0<< \sigma$, 
the Enskog equation is inadequate for describing spatial variations 
with wavelength $\lambda$ in the range $\ell_0 < \lambda < \sigma$. 
These wavelengths are relevant for analyzing dynamic structure 
factors as measured in neutron scattering experiments. However, the revised 
Enskog equation of section 12 is very adequate.}, though approximate, generalization of the Boltzmann equation, to 
account for 
transport processes in a dense fluid of hard spheres. This was 
achieved by replacing
Boltzmann's molecular chaos assumption (2) by
\be
f^{(2)}(x_1 x_2,t) \lra g( | \vr_{12} | = \sigma ) f(\vr_1, \vv_1 , 
t)
f(\vr_1 - \mbox{\boldmath$\sigma$}, \vv_2 , t) .
\ee
The local equilibrium function for two spheres at contact, $g(\sigma 
)$,
accounts for static correlations in a hard sphere fluid, and the 
different
space dependence of the two $f$'s accounts for the difference in 
position
of the two colliding spheres, and represents the new transport 
mechanism of
{\it collisional transfer}, i.e. instantaneous transfer of momentum 
and energy
over the hard sphere diameter $\sigma$ through the interaction. 
This is in general the dominant
transport mechanism at high densities. The original Boltzmann 
equation takes
only {\em kinetic transport} into account. 

\section{BBGKY-hierarchy}
The {\it Renaissance} began in 1946 with
Bogoliubov's article {\it Problems of a dynamical theory in
statistical physics} \cite{Bogol}. The fundamental basis was the 
BBGKY-hierarchy \cite{Coh-Nij}, which is named after Bogoliubov, 
Born, Green, Kirkwood and Yvon, and reads 
\bea
(\partial_t + \vv_1 \cdot \mbox{\boldmath $\nabla$}_1)f_1 & = & \int 
dx_2  
\theta_{12} f^{(2)}_{12} , \nonumber \\
(\partial_t + {\cal L} (12)) f^{(2)}_{12} & = & \int dx_3 
(\theta_{13} +
\theta_{23} ) f^{(3)}_{123}
\eea
and so on,
where $\theta_{12}$ is the weak scattering operator,
\be
\theta_{12} = \frac{\partial V (r_{12})}{\partial \vr_{12}} \cdot 
\frac{1}{m}
\left( \frac{\partial}{\partial \vv_1} - \frac{\partial}{\partial 
\vv_2}
\right) ,
\ee
with pair potential $V(r)$, and two-particle Liouville operator
\be
{\cal L}(12)  =  {\cal L}^0 (12) - \theta_{12} 
 \equiv  \vv_1 \cdot \mbox{\boldmath $\nabla$}_1 + \vv_2 \cdot 
\mbox{\boldmath$\nabla$}_2 - \theta_{12}.
\ee
This set of equations is an open hierarchy of coupled equations that 
expresses
the time rate of change of the $s$-particle distribution function 
$f^{(s)}$
in terms of the $(s+1)$-particle distribution function, provided the
interactions are pairwise additive. 

Bogoliubov and Kirkwood were
interested in non-equilibrium properties of fluids and dense 
gases~\cite{Brush}.
Uhlenbeck considered Bogoliubov as the originator of modern methods 
in
kinetic theory, as he introduced the basic concepts of systematically
approximating the dynamics of moderately dense gases in terms of 
small groups
of two, three, four,... isolated particles, just like in Mayer's 
cluster
expansion for the equation of state in thermal 
equilibrium.
This set the stage in 1954 for Choh's thesis work.

The main goal of this paper is to analyze the impact of the
Bogoliubov Choh Uhlenbeck theory on later developments. Consequently 
I will refer to links with later
developments in kinetic theory for Lattice Gas Automata~\footnote{
LGA are fully discretized models for non-equilibrium fluids \cite{6-gang},
defined on a regular space lattice, where time is discrete, where point 
particles live on the nodes of a lattice, and have a small set of allowed
velocities corresponding to the $b$ nearest neighbor lattice vectors. The
dynamics consists of a collision step, followed by a propagation step. If
$\ell$ particles ($\ell = 2,3,\ldots ,b)$ are on the same node (zero range
interactions), then they suffer $\ell$--tuple collisions according to
well defined collision rules, which are obeying momentum conservation
in fluid models, and lacking it in purely diffusive or reactive models.
 As the interactions in LGA are not
pairwise additive, but involve genuine $\ell$--tuple collisions, the right
hand side of the first hierarchy equation for LGA involves already 
all distribution
functions $f^{(\ell )}$ with $\ell = 2,3,\ldots , b+1$.
Ofcourse, a similar structure occurs in continuous fluids with non-additive 
$b$--body interactions.} (LGA),
in which the BBGKY-hierarchy has been used as well~\cite{Buss+Dufty}. 

An extremely important parallel development occurred also in 1954, 
which 
led to the Green Kubo formulas or {\it time correlation function} 
method
for transport coefficients in fluids \cite{Brush}. The shear viscosity, for 
instance, is given by
\be
\eta = \frac{1}{Vk_B T} \int^{\infty}_{0} dt \langle J_{xy}(t) J_{xy} 
(0) \rangle ,
\ee
where $J_{xy}(t)$ is the microscopic $N$-particle momentum flux, which 
contains both
kinetic as well as collisional transfer contributions. The generic 
form of a
Green Kubo formula is a time integral over an {\em equilibrium} time
correlation function, such as the stress correlation function in (7).
 It involves in fact two limits: first the 
thermodynamic
limit of $\langle \ldots \rangle /V$ is taken, followed by a time 
integral
over an infinite time interval \cite{Zwanzig-K,McLennan-K}.
Similar Green Kubo formulas for LGA were derived by Dufty and Ernst
 \cite{Kubo-LGA}, in which  the time integrals are replaced by time sums
and  the initial term in the summand has a weight one half.

\section{Bogoliubov Choh Uhlenbeck theory}
This section will only give an outline of the basic ideas of
this theory. The interested reader is referred to Cohen's lectures~\cite{Coh-Nij} about this subject.

The system of interest is a moderately dense gas under conditions of 
standard
temperature and pressure, composed of particles interacting through 
short 
range, strictly repulsive forces. Here the typical duration of a 
collision is
$\tau_0 = 10^{-12}$ seconds, whereas the mean free time is $t_0 = 
10^{-9}$
seconds. The new concept is that the existence of well separated time 
scales
brings about a contraction of the description in terms of fewer 
variables, as
time progresses.

There exist three time stages: initial stage, kinetic stage and 
hydrodynamic stage. In the {\it initial stage} $(t < \tau_0 )$ the 
full
Liouville equation is required to describe the fast variation of the 
$N$-particle
distribution function. In the subsequent {\it kinetic stage} $(\tau_0 
< t < t_0 )$,
which is the first coarse grained time scale, the single particle 
distribution
function $f(x,t)$ is the only slow variable, and the higher order 
function
$f^{(s)}$ are assumed to be {\it time independent functionals}
$f^{(s)}(\ldots | f(t))$ of $f(x,t)$, which implies a loss of memory 
about the initial state~\footnote{As will be discussed later on, it is the assumption about 
the
existence of a time-independent functional, to general order in the 
density,
that breaks down and leads to the divergence of the density 
expansion of transport coefficients, to be discussed in section 5.}.
Once $f^{(2)}(x,x_2 | f(t))$ is
determined, one has a closed kinetic equation, the generalized 
Boltzmann
equation. 
The perturbative calculation proceeds by performing a formal 
{\it density expansion}.
In the final {\it hydrodynamic stage} $(t > t_0)$, which is the second coarse
grained time scale, a further contraction of the description occurs. 
Here
the only slow variables are the hydrodynamic fields $a(\vr ,t) = \{ 
n(\vr , t),
\vu (\vr ,t), T(\vr ,t) \}$ and $f(x,t)$ becomes a {\it time}- and
{\it space}-independent functional of these fields, $f(\vr ,t) =
f(\vv | a(\vr ,t))$. This assumption is a generalization of the 
normal
solution in the Chapman Enskog method. Once $f(\vv | a (\vr ,t))$ has 
been
determined, one has obtained a closed set of hydrodynamic equations, 
which
contains the Navier Stokes equations, and generalizations thereof, as 
well as
explicit expressions for the transport coefficients.
The perturbative calculation proceeds by performing a formal
$\mu$--expansion, where $\mu \sim \ell_0 \nabla$ is the non-uniformity parameter.

How can these ideas be implemented to actually derive the generalized
Boltzmann equation? Clearly, in order to solve an open hierarchy, 
like (3),
some kind of closure relation is required. In the present theory 
closure is 
obtained by imposing the {\it factorization assumption} as a boundary 
condition
on the BBGKY-hierarchy at $t= - \infty$, expressing the absence of
correlations in the infinite past. This is technically expressed 
through
the relation
\[
\lim_{\tau\ra\infty} S_{-\tau} (12) f^{(2)} (x_1 x_2| 
S^{0}_{-\tau} f(t)) = 
\]
\be
{\cal S} (12) f(x_1 ,t) f(x_2,t) \equiv f(x^{*}_{1},t) f(x^{*}_{2},t) ,
\ee

\noindent
and similar relations for higher order distribution functions. Here
$S_t (12)$ is the 
{\em streaming} operator for two interacting particles, $S_t (12) = 
\exp [t{\cal L}(12)]$, and $S^{0}_{t}(12)$ is the {\it free streaming} 
operator,
$S^{0}_{t}(12) = \exp [t{\cal L}_0 (12)]$, where the Liouville 
operators
${\cal L}$ and ${\cal L}_0$ are defined in (6).
The streaming operators $S_t (12)$ and $S^{0}_t (12)$ generate the trajectories of two isolated particles with or without interaction 
respectively.
 Moreover ${\cal S}(12)$ is defined through
\be
{\cal S}(12) = \lim_{\tau\ra\infty} S_{-\tau}(12) S^{0}_{\tau}(12) .
\ee
Both streaming operators ${\cal S}(12)$ and $S_{-t}(12)$ act 
directly on the phases $x_i$ with $i=1,2$, by replacing 
them respectively by 
$x^{*}_{i} =
{\cal S}(12)x_i$ and $x_i(-t) = S_{-t}(12)x_i$. The latter operator 
generates backward dynamics. The action of both streaming operators 
on $x_1$ and $x_2$ is sketched in  Figure 1.

\begin{figure}[t]
\centerline{\psfig{file=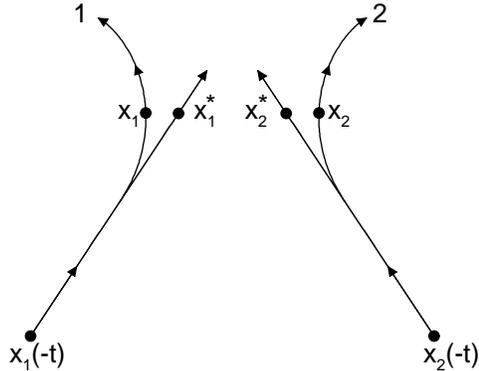,height=5cm}}
\caption{Action of streaming operators.}
\end{figure}

It is exactly the factorized boundary condition at $t = -\infty$, 
which
introduces the {\it arrow of time} into the Boltzmann equation and 
its
generalizations. The BBGKY-hierarchy equations themselves -- without 
the
boundary condition -- are invariant under  time reversal, as required by 
classical mechanics. The final ingredient
in obtaining the generalized Boltzmann equation is the use of the small 
parameter
density, $n=N/V$, as a systematic expansion parameter. It enables 
one to
perform an $n$-expansion of the unknown functionals $f^{(s)} = 
f^{(s)}_{0} +
nf^{(s)}_{1} + \ldots$, of the equations of the BBGKY-hierarchy, 
and of the
boundary condition (8). 

\section{Choh Uhlenbeck equation}
The implementation of these ideas were realized in Choh's thesis, and 
led to the {\it Choh Uhlenbeck} equation in 1958: 

\be
(\partial_t + \vv \cdot \mbox{\boldmath $\nabla$}) f = I(ff) + K(fff) 
+ \ldots
\ee
where $I(ff)$ is the binary collision term,
\be
I(ff) = \int dx_2 \theta_{12} {\cal S}(12) f(x_1,t) f(x_2,t) ,
\ee
and $K(fff)$ is the triple collision term, which will be specified 
below. It is  an important new
result, that generalizes the Boltzmann equation to include triple
collisions. 
Bogoliubov had already shown in his 1945 article that $I(ff)$ for the 
case of
hard spheres reduces to the collision term in the 
Enskog Boltzmann equation with $g(\vr_{12} | = 
\sigma ) = 1$.
This binary collision term $I(ff)$ does take into account the 
difference in
position of the colliding particles, the so-called collisional 
transfer.
In the limit of low density, it gives a higher order density 
correction to
the Boltzmann equation.

In the notation of Choh's thesis, the triple collision operator has 
the form,
\be
K(fff) =
\int dx_2 \theta_{12} \int^{\infty}_{0} d\tau S_{-\tau}(12) 
\int dx_3
[ \ldots ] S^{0}_{\tau}(12) \Pi^{3}_{i=1} f(x_i,t)
\ee
with
\be
[\ldots ] = \int dx_3 [ ( \theta_{13} + \theta_{23}) {\cal S}(123) 
- {\cal S}(12) \theta_{13} {\cal S}(13) - {\cal S} (12) 
\theta_{23} {\cal S} (23) ] .
\ee
A more transparent form was obtained in the later cluster expansion 
methods of
Green and Cohen (see review \cite{DvB}),
namely
\bea
K(fff) & = & \int dx_2 \theta_{12} \int dx_3  \{ {\cal S} (123) - {\cal 
S}(12)
{\cal S}(13) \nonumber \\
& - & {\cal S}(12) {\cal S}(23) + {\cal S}(12)\} f(x_1) f(x_2) f(x_3) .
\eea
It is instructive to analyze the different types of contributions from 
${\cal S}(123)$, illustrated in Figure 2. 
\begin{figure}
\centerline{\psfig{file=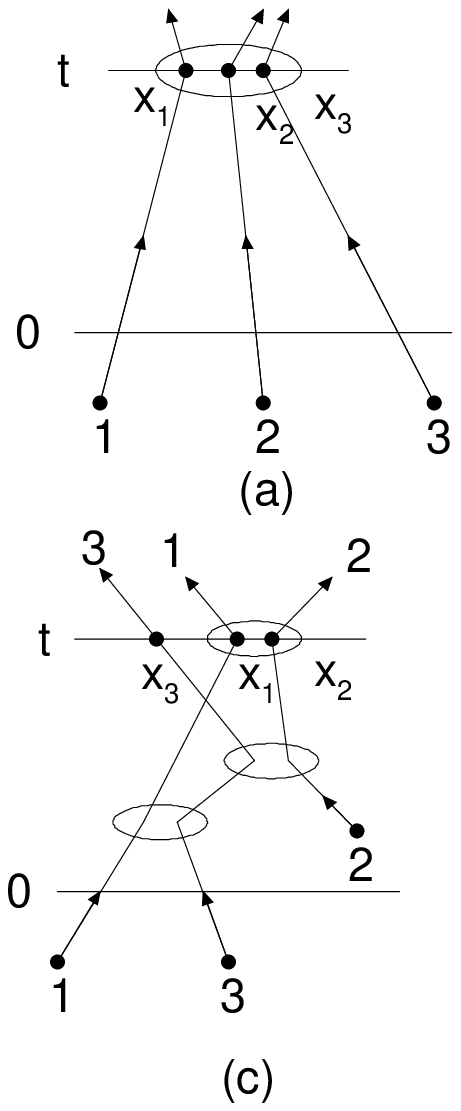,height=7cm} \hspace*{1.2cm}
\psfig{file=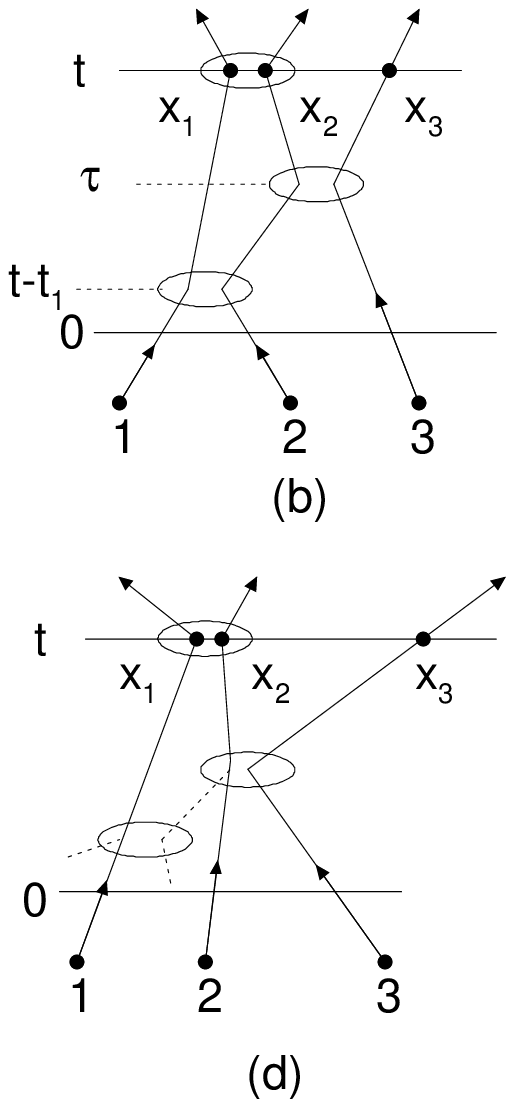,height=7cm}}
\caption{ Genuine triple collisions (a) and ring collisions: 
 recollision (b), cyclic (c) and hypothetical 
(d) collision.}
\end{figure}
They contain {\it genuine} triple collisions, where 
three
particles are simultaneously in each others interaction sphere. Once 
they
have collided, the particles will separate and never recollide again due to 
the short
range repulsive potential. In hard sphere systems, where interactions 
are
instantaneous, this term is vanishingly small.

Moreover, ${\cal S}$(123) contains sequences of two {\it uncorrelated} 
successive
binary collisions (12) (13) and (12) (23). Such collisions have 
already
been accounted for in the Boltzmann collision term $I(ff)$. So, they 
are
subtracted out from the possible sequences of binary collisions among 
three
particles in (14).

Finally ${\cal S}$(123) contains sequences of three {\it correlated }
successive
binary collisions, such as the recollision (12)(13)(12) and the 
cyclic
collision (12)(23)(13), generally referred to as {\em ring collisions}. 
A more systematic analysis of the
separate terms may be found in Ref \cite{DvB}.

In the second part of his thesis, Choh obtained the solution to the 
Choh Uhlenbeck equation
in the hydrodynamic stage by following Bogoliubov's generalization 
of the
Chapman Enskog method, and derived formal expressions for the first 
density
correction to the viscosity and heat conductivity for a moderately 
dense gas in three dimensions with short range repulsive interactions.
The  viscosity, for instance, has the form
\be
\eta = \eta_0 + \eta_1 n + \eta_2 n^2 + \ldots,
\ee
where $\eta_0$ is the low density Boltzmann result and $\eta_1$ is 
the Choh Uhlenbeck contribution. It is expressed
in terms of matrix elements of the triple collision operator, and
contains
contributions from genuine triple collisions, as well as from the 
correlated
ring collisions, illustrated in Figure 2. For the special case of hard 
spheres,
Sengers and coworkers have computed
$\eta_1$, and found that $\eta_1$ differs only a few percent from the
predictions of the Enskog equation~\cite{Coh-Berl}. 

Similar calculations for the viscosity in LGA~\footnote{
No reviews on the kinetic theory beyond the Boltzmann mean field equation 
are available for LGA. Therefore I refer in each section to the original 
LGA publications to show the parallel developments.}
 have been carried 
out by Brito et al.~\cite{Matsui}. Here it was found that the ring 
collisions give corrections on the order of 10 \%  to the
 Boltzmann mean field results for {\it all} densities, and the 
resulting viscosities are in excellent agreement with those of 
computer simulations. I want to stress that the Boltzmann mean field
equation and the ring kinetic equation for LGA are {\it not} 
restricted to low densities. The Boltzmann equation for LGA does include
apart from binary collisions {\it genuine} triple, quadruple, 
$\cdots, b$--tuple 
collisions, as well as static correlations, accounting 
for the exclusion principle, which mimics the hard core interactions.

The seven years after the appearance of
Choh's thesis in 1958 may be described as a period of consolidation. 
Using
the cluster expansion method~\cite{DvB}, Green and Cohen derived the formal 
structure
of the $\ell$-tuple collision terms $(\ell = 2,3,4,\ldots )$ in the
generalized Boltzmann equation, in close analogy to the Mayer cluster 
expansion for reduced distribution functions in thermal equilibrium, 
and it
was believed that transport coefficients could be calculated in the 
form of
a virial expansion where  the coefficients $\eta_\ell$ in (15) 
were
determined by the dynamics of $(\ell +2)$ isolated particles.
Similar cluster expansion methods were applied to the Green Kubo 
formulas
for viscosity and heat conductivity. After some initial confusion 
about a
possible difference between the Green Kubo formulas and the 
Choh Uhlenbeck 
equation, McLennan was the first to show that both methods yield the 
same
results for the first density correction to the transport 
coefficients in fluids \cite{McLennan-Lett}.

\section{Discovery of the divergence}
The cluster expansion method of Green and Cohen or the alternative 
inversion method of Zwanzig \cite{Zwanzig} made it feasible 
to investigate the fundamental assumption of the 
Bogoliubov Choh Uhlenbeck theory about the rapid decay of initial 
correlations within a time $t_o$. This assumption provided the formal justification for the rapid approach
of the pair- and higher order distribution functions $F^{(s)}(x_1 x_2 
\ldots x_3 ,t)$ to the {\it time independent} functionals 
$f^{(s)}(x_1 x_2 \ldots x_3| f(t))$.

In 1965 Dorfman and Cohen, Weinstock, as well as Goldman and Frieman 
were the first to make phase space estimates of the triple and higher
 order collision terms \cite{Brush}, and to show that the virial 
expansion contains coefficients that are {\em divergent} with increasing time; more specifically
\be
\eta = \left\{ \begin{array}{ll}
\eta_0 + \eta_1 n + \eta_2 (t) n^2 + \ldots \quad & (d=3) \\
\eta_0 + \eta_1 (t) n + \ldots \quad & (d=2)
\end{array} \right. 
\ee
where the quadruple collision contribution $\eta_2 (t)$ in three 
dimensions  and the
triple collision contribution $\eta_1 (t)$ in two dimensions are 
logarithmically
divergent as $t \to \infty$~\cite{Brush}. Kawasaki and Oppenheim were 
the first to sum the ring diagrams. Subsequently the divergences 
and resummations were consolidated by more detailed calculations of 
van Leeuwen and  Weijland, Swenson and McLennan, Haines together 
with Dorfman and 
Ernst, Sengers, and Fujita~\cite{Brush,EHD}.

It is of interest to elaborate on this fundamental discovery, which 
marks
the advent of the {\it Modern Era} in kinetic theory, where 
collective 
effects determine the behavior of transport coefficients, even at low densities. To do so, I
present the derivation of the $\log t$ divergence, as given in Ref.~\cite{DvB}
for hard sphere models and Lorentz gases.

Consider the $x_3$-integral in (14) at fixed $x_1 , x_2$. Let 
$\int^{*} dx_3$
denote the phase space where the integrand of the recollision term 
(12)(13)(12) is non-vanishing, and let $\sigma$ be the range of the repulsive 
interaction.
The configuration with $dx_3 = d\vv_3 d\vr_3$ for the occurrence of a
recollision can be simply estimated in the coordinate system used in 
Figure 3. 

\begin{figure}[h]
\centerline{\psfig{file=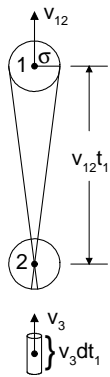,height=5cm}}
\caption{Phase space for recollisions.}
\end{figure}

The solid angle $d\hat{\vv}_3$, in which particle 3 has to hit 
particle 2,
such that particle 2 will recollide with particle 1 after a time 
$t_1$, is proportional to $d\hat{\vv}_3 \sim (a/v_{12} t_1)^{d-1}$. For 
particle 3 to
perform this well-aimed collision its position has to be inside a 
small
collision cylinder $d\vr_3 \sim a^{d-1} v_3 dt_1$. The total phase 
space for 
the recollision event is then obtained by integrating over all 
possible
intermediate times, yielding for large times $t$:
\be
\int^* dx_3 \sim \int^{t}_{\tau_0} dt_1 \left( \frac{a}{v_{12} t_1} 
\right)^{d-1} \sim
\left\{ \begin{array}{cl}
\log t  \quad & (d=2) \\
{\rm const} + t^{2-d} \quad & (d \neq 2)
\end{array} \right.
\ee
This estimate illustrates that the triple collision term in the
Choh Uhlenbeck equation in three-dimensional systems approaches a finite 
limit,  implying that $\eta_1$ in (16) exists, and that the initial state
is  indeed forgotten, in agreement with Bogoliubov's 
assumption
about the existence of a time-independent functional.
However all three conclusions, valid for three dimensions, break down in two
dimensions, where the Choh Uhlenbeck term diverges as $\log t$.

Next, I consider a similar phase space estimate for the ring 
collisions
in the quadruple collision term, for instance (12)(23)(24)(12).
The additional $x_4$-integration yields an extra factor $t_1$ in (17). Consequently the relevant phase space,
\be
\int^{*} dx_3 \int^{*} dx_4 \sim  \int^{t}_{\tau_0} dt_1 t_1 t^{1-d}_{1} \sim
\left\{ \begin{array}{cl}
\log t \quad & (d=3) \\
{\rm const} + t^{3-d} \quad & (d\neq 3) ,
\end{array} \right.
\ee
{\it diverges logarithmically} in three dimensions as $t \to \infty$, and a virial 
expansion of 
transport coefficients does not exist in three dimensions either. 
Therefore, the memory
of the {\it initial} state is not completely lost. 

The above estimates show that dynamic correlations create long time
correlations so that the initial and kinetic stage are {\it not} 
well {\it separated} anymore. This observation invalidates the first part of Bogoliubov's assumption about the existence of well separated time scales.
These long time correlations are created by
unrestricted free motion, which is possible in isolated groups of 3, 
4, ...
particles. However infinitely long free paths between collisions are
unphysical in a real gas. The free motion of a particle will on 
average 
be on the order of the mean free path $l_0 \sim \overline{v} t_0 \sim 
1/n$.
Therefore, the {\it collective effect} of the other particles, i.e. 
the
surrounding fluid, provides an effective cut off which replaces $t$ in 
 (17) and (18) by $t_0$. Consequently, the transport coefficients 
for
moderately dense gases exhibit a non-analytic term
$n \log t_0 \sim n | \log n |$ in the density, i.e.
\be
\eta = \left\{ \begin{array}{ll}
\eta_0 + \eta_1 n + \eta^{\prime}_{2}n^2 \log n \quad & (d=3) \\
\eta_0 + \eta^{\prime}_{1}  \log n \quad & (d=2)
\end{array} \right.
\ee
where the coefficients $\eta^{\prime}_{1}$ and $\eta^{\prime}_{2}$
have been calculated for hard sphere fluids~\cite{DvB} and 
Lorentz~\footnote{The Lorentz gas is a simple model for diffusion 
and consists of independent light particles moving through a random
 array of fixed scatterers.} gases~\cite{Sitges}.

The analytical results (19) for transport coefficients can be considered 
as well established for fluids with short range repulsive interactions, 
and are believed to be valid for realistic interparticle potentials as well.
The low density expansion (19) appears to be consistent with 
experiment~\cite{experim}, but so is the virial expansion (15). However,
 up to the present time there are no compelling direct experimental
 observations for the appearance of a logarithmic term  $n^2 \log n$ 
in the density expansion of the transport coefficients.

\section{Necessity of log ${\bf n}$ singularities?}
The divergence of the ring collision integrals, discussed in (17) and (18), 
holds for {\it continuous} velocity variables, and so does the logarithmic
singularity (19) in the
low density limit. However in systems with {\it discrete} velocity 
directions, such as the 
Ehrenfest wind-tree model \cite{Sitges}, $d$-dimensional LGA \cite{Binder},
and even one-dimensional continuous velocity models \cite{Jepsen}, the ring
collision integrals diverge algebraically, instead of logarithmically.
This can have very surprising effects.

Consider, for instance, the self diffusion coefficient $D$ in a
one-dimensional gas of hard rods, which has been calculated 
exactly \cite{Jepsen}. Calculating  $D$ from a formal density expansion 
yields $nD = d_0 + d_1 (t) n+ \ldots$, where $d_1(t) \sim t$ for large
$t$. By taking the collective effects of the surrounding fluid into account,
$t$ is effectively replaced by the mean free time $t_0 \sim 1/n$. The
renormalized low density behavior of the diffusion coefficient becomes 
$nD = (d_0 + d^{\prime}_{1})$ for small $n$, i.e. the ring collisions modify
the leading order Boltzmann contribution. Exactly the same renormalization
of the Boltzmann diffusion coefficient through ring collisions has been found
in stochastic lattice Lorentz gases with back scattering \cite{Binder}. 
The exact 
low density diffusion coefficient for these models, which is very 
different from the
Boltzmann value, has been calculated by van Beijeren and Ernst \cite{vBE}
through an exact enumeration method. 

These two examples have been chosen in order to illustrate (i) that the
behavior of transport coefficients, even at  low or moderate  densities, is 
not necessarily determined by the dynamics of small isolated groups of
2,3,4,... particles, as had been assumed in the Bogoliubov Choh Uhlenbeck
theory, but that it is strongly influenced by {\it collective effects},
and (ii) that the ring collisions {\it not} necessarily lead to (logarithmic)
{\it singularities} at small densities. It all depends
on the structure of the relevant phase space.

The previous discussion may suggest that in principle the transport
properties in two- and three-dimensional systems are well understood.  We know
the coefficient of the dominant singularity $n^{d-1} \log n$ in $d$
dimensions $(d=2,3)$. We know how to renormalize all divergent terms 
of the $n$--expansion into finite contributions by resumming the most 
divergent ring diagrams.
It looks as if only technical difficulties are stalling further progress.

However, we do not know the total coefficient of ${\cal O}(n^{d-1})$. It was Pomeau~\cite{Pomeau}, who first realized that the total coefficient of the 
${\cal O}(n^{2})$--term in two dimensions was {\it divergent} as well, and 
that there was something fundamentally wrong with the Navier Stokes 
equations and  transport coefficients in two-dimensional fluids, whereas the
hydrodynamic equations in three dimensions are still well defined to 
Navier Stokes order.

To conclude this section, it is worth mentioning that  
calculation of terms beyond the dominant $n^2 \log n$--term in 
three dimensions, for instance the ${\cal O}(n^2)$--contribution, 
 is exceedingly 
difficult, and general results are lacking. In fact, no progress along these 
lines has been made since the
early eighties. The systematic kinetic theory for fluids has reached an 
impasse, and the high goals of the modern era have unfortunately not been realized. The reader, interested in further details, is referred to the 
 review by Dorfman and van Beijeren~\cite{DvB}. Although twenty 
years old, this review still gives the best account of the current status of the systematic kinetic theory.

\section{Long time tails}
In 1970 Alder and Wainwright \cite{Alder} made their famous discovery
of the long time tails in the velocity  correlation and other current
correlation functions in
{\it equilibrium fluids}, although a first indication of the existence of 
 long time tails in three dimensions was already present in a paper of 1966 by 
Goldman~\cite{Goldman}.
Computer simulations on hard sphere 
systems \cite{Alder,Wood} show that the velocity correlation function (VAF)
 of a tagged particle, $\langle v_x (t) v_x (0) \rangle$, has a {\it 
positive}
long-time tail, proportional to $t^{-d/2}$, which is caused by the following
mechanism \cite{Alder}. The initial momentum of the tagged particle is
transferred in part to the surrounding fluid particles. This sets up a flow
 of vortices around the tagged particle, which transfers some of its
momentum back to the tagged particle (``kick in the back''), thus yielding a 
positive correlation between $\vv (0)$ and $\vv (t) $. This correlation extends
over hydrodynamic time scales. In LGA similar long time tails have been 
observed with much higher statistical precision and over much longer time
intervals, using Frenkel's moment propagation method \cite{Frenkel}.
The long time tails are a convincing illustration of the important 
role that computer simulations have played in unraveling the complex 
dynamics of classical fluids.

The hydrodynamic picture can be cast into a more quantitative form by using 
the mode coupling theories, developed for critical dynamics~\cite{MC-crit,Pom+R},
to fluids away from  crytical points. As an example consider $\langle v_x(t) v_x(0) \rangle$ 
for a tagged particle. To describe the long time behavior
on a coarse grained spatial and temporal scale,
 I decompose the total tagged particle current,
${\bf v}(t) \simeq \int  d{\bf r} n_s({\bf r},t) {\bf u}({\bf r},t)$ in 
terms of the two slow modes: tagged particle density $n_s({\bf r},t)$
and fluid flow velocity ${\bf u}({\bf r},t)$. The relevant Fourier 
modes decay as $\hat{n}_s({\bf k},t) \simeq \exp (-Dk^2t) \hat{n}_s({\bf k},0)$ and  $\hat{{\bf u}}({\bf k},t) \simeq \exp (-\nu k^2t) 
\hat{{\bf u}}({\bf k},0)$, where $D$ is the coefficient of self
 diffusion and $\nu = \eta/\rho$ the kinematic viscosity. Consequently, the 
total current decays as 
\be
v_x (t) \sim  \int \frac{d{\bf k}}{(2 \pi)^{d}}\exp[-(D+\nu)k^2t] 
\sim [4 \pi (D+\nu)t]^{-d/2}. 
\ee
This leads to the quantitative explanation of long time tails, as given by 
the author, Hauge and van Leeuwen~\cite{Trondheim}. 
A more fundamental approach through resummation of ring
collision events was given by Dorfman and Cohen \cite{Bob-tails}. The former theories 
are valid for general densities, the
latter are restricted to low and moderate densities. 
I also want to point out that the kinetic theory
approach of Pomeau~\cite{Pomeau}, together with that of Dorfman and 
Cohen, in which the ideas of Pomeau's 1968 articles were incorporated 
and clarified, might be considered as the origin of ring kinetic theory, 
as it is used at the present times. 

The predicted {\it asymptotic} long time tails in the stress correlation function in (7) and in the other Green Kubo integrands have never been
 observed in computer simulations in hard core fluids. They are 
overshadowed by the molasses tails, which are present on 
{\it intermediate} time scales, and are related to the cage 
effects ( see section 11). However in LGA with only on-node interactions the mechanism for "building cages from surrounding hard objects" is absent, 
and the observed long time tails agree reasonably well with theoretical 
predictions~\cite{Naitoh}.

In purely diffusive models, such as Lorentz gases \cite{Sitges},
the correlation function decays as $\langle v_x (t) v_x (0) \rangle 
\sim -t^{-1-d/2}$, i.e.
after a long time $t$ the moving particle returns to its point of origin
on average moving in a direction opposite to its initial velocity, instead
of parallel as in the fluid case. Hence, the correlations are {\it negative}
at large times and weaker than in fluids.
In continuous Lorentz gases, as well as in LGA the observed long time 
tail phenomena were quantitatively explained by means of mode
coupling theory \cite{Machta,Frenkel} as well as by ring
kinetic theory \cite{Sitges,Brito}. No new fundamental 
problems seem to arise here.

Experimental confirmation of the long time tails for air and gaseous argon has 
been obtained by Kim and Matta \cite{Kim-Matta}, and for latex spheres in water by Ohbayashi et al.~\cite{japanese} using the photon 
correlation mehtod. Neutron scattering
experiments on liquid argon and sodium also offer indirect evidence for the
existence of long time tails (for references see \cite{Coh-Berl}).

\section{Renormalized fluid dynamic equations}
The most dramatic consequence of the long time tails in fluids 
is the non-existence
of the Navier Stokes equations in two dimensions \cite{Pomeau}. 
As the {\it transport
coefficients} of viscosity, heat conductivity and self-diffusion  
 are expressed through  Green Kubo formulas as time integral over current
correlation functions, these time integrals are diverging as $\log t$, and
transport coefficients do {\it not exist} in two-dimensional fluids. In
purely diffusive systems, like Lorentz gases, on the other hand, Fick's law is
still valid in two-dimensions, since $\langle v_x (0) v_x(t) \rangle \sim
1/t^2$ for large $t$, and the diffusion coefficient $D$ is finite.

In three dimensions conventional hydrodynamics involves constitutive 
relations where (Fourier  transforms of) fluxes are linear 
in the gradients or in the wave number ${\bf k}$ and  the coefficients
 of proportionality are the Navier Stokes transport coefficients. 
In the same conventional picture one expects that the fluxes can be 
expanded systematically in powers of $k$, where the coefficients of
 proportionality in the ${\cal O}(k^2)$--terms are the  Burnett 
coefficients. This systematic expansion in the {\it non-uniformity 
parameter}, $\mu \sim \ell_0 \nabla$, was the basis of Bogoliubov's 
generalization of the Chapman Enskog method in section 3.

However, explicit calculations using mode coupling theory or ring 
kinetic theory \cite{Dufty-B,DE-Burnett} show that the 
Burnett coefficients in three dimensions are diverging 
logarithmically. This is a consequence of the occurrence of algebraic long time tails, which show that the kinetic and hydrodynamic time stages are not well separated. This  observation invalidates the second part of Bogoliubov's assumption about the existence of well separated time scales.

We observe that there is in fact a close parallel between on the one 
hand the non-existence of the {\it density expansion} in (16), 
which resulted in a non-analyticity in the low density behavior 
of the transport coefficients, and on the other hand the 
non-existence of the $\mu$--expansion, {\it i.e.} the non-existence  of the Navier Stokes equations in two dimensions, and the non-existence of the 
Burnett equations in three dimensions.

The legitimate question is then: Are there any singularities 
in the renormalized $\mu$--expansion?
If the Navier Stokes equations 
do not exist in two dimensions, what is the structure of the 
renormalized hydrodynamic equations in two-dimensional fluids? 
Similarly, what is the structure of the renomalized fluid 
dynamic equations in three dimensions? Mode coupling theory \cite{Pomeau2,5-gang,Pom+R,DE-Burnett}, as well as kinetic 
theory \cite{Ring-ED} provide  partial answers.

Suppose one calculates the dispersion relation for the relaxation 
rate $z_{\perp}(k)$ of transverse velocity excitations (shear modes). 
Conventional hydrodynamics gives $z_{\perp}(k) = - \nu k^2 + \cdots$, 
where $\nu = \eta/\rho$ is the kinematic viscosity. 
The new theory yields:
\be
z_{\perp}(k)= \left\{ \begin{array}{ll}
- \nu k^2 + \nu_1^\prime k^{5/2} + \ldots \quad & (d=3) \\
- \nu^\prime k^2 \log k + \ldots \quad & (d=2)
\end{array} \right.
\ee
The small-$k$ singularity in the dispersion relation suggests that 
the constitutive relations between fluxes and gradient fields 
involve transport kernels which are nonlocal in space and time 
\cite{Dufty-B,Alder-Today}. There exist also more complex systems with
intrinsic length scales larger than $\ell_0$ ( e.g. dense fluids ) where generalized hydrodynamics with wavenumber dependent elastic and dissipative reponse functions are relevant~\cite{Coh-Berl}. The same holds for LGA~\cite{Boon}.
 
The general structure of the renormalized fluid 
dynamic equations in two and three dimensions is still an open problem. 
A clear resolution of their structure is hampered by the apparently
 complicated non-local structures produced by the long range 
spatial and temporal correlations, as well as by the {\it smallness} 
of the effects that these correlations produce when they are 
calculated or hunted for experimentally. As the effects of long time tails are
very small in three dimensions, the use of the conventional Navier Stokes equations should ofcourse continue in the forseeable future.

\section{Long range correlations in nonequilibrium stationary states}
The first author to analyze already in 1978 {\it long range spatial correlations} in 
non-equilibrium fluids in terms of ring collision events was 
Onuki \cite{Onuki}. He calculated the pair distribution function in
a steady laminar flow using the Boltzmann-Langevin equation, and showed 
that the pair correlation function has a very long range, exhibiting Coulomb-type behavior, and is proportional to log $r$ for $d=2$ and to 
$1/r$ for $d=3$. Kirkpatrick et al. (see review~\cite{Coh-Berl,DKS})
 used in 1982 ring kinetic theory to calculate long rang correlations 
in fluids with an imposed temperature gradient (here the range is even
 longer than in 
the laminar flow case), and calculated the dynamic strucure factor, 
which is measured in light scattering experiments.

Such spatial correlations occur in {\it open systems} that
are kept out of equilibrium by  reservoirs \cite{DKS} or by external driving fields \cite{Zia}. The objects of interest here are the
{\it spatial} correlations between fluctuations in the local density 
$\delta n(\vr ,t) =  n(\vr ,t) - \langle n \rangle$ or in the local flow
velocity $\delta\vu (\vr ,t) = \vu (r,t) - \langle u \rangle$. The
correlation functions are taken at {\it equal} or at {\it unequal} times.
The spatial Fourier transforms of the former are the static structure 
factors; the spatial and temporal Fourier transforms of the latter are the
dynamic structure factors. The correlations and structure factors related
to density fluctuations are studied in light and neutron scattering
experiments \cite{DKS,Coh-Berl}; those related to fluctuations in 
the flow field are
studied in homogeneous turbulence \cite{Batchelor}. The theoretical
and experimental results for non-equilibrium stationary states in fluids 
with externally imposed gradients in temperature or flow field have been
recently reviewed in \cite{DKS}. Those for driven diffusive 
systems are discussed in a recent monograph by Schmittmann and Zia \cite{Zia}.

Also here it is instructive to discuss a simple example: I consider the density-density correlation function $G({\bf r})$ and corresponding susceptibility or structure factor 
$\chi({\bf k})$ in the stationary state of a spatially uniform driven 
system, where the driving field violates the conditions of detailed 
balance and breaks the spatial isotropy, {\it i.e.}
\bea
G({\bf r}) &=& \frac{1}{V} \int d{\bf r}^\prime   
\langle \delta n({\bf r}^\prime
 +{\bf r}) \delta n({\bf r}^\prime ) \rangle \nonumber \\
&=&  \frac{1}{(2 \pi)^d} \int d{\bf k} \chi({\bf k}) \exp(i{\bf k\cdot r}).
\eea
Clearly, for the correlation function $G({\bf r}) $ to have an algebraic 
tail, the susceptibility $\chi({\bf k})$ needs to have a singularity 
at small wave number ${\bf k}$. The singularities discussed in 
section 8 are {\it branch points} of the form log $k$ or $k^\alpha$ with non-integer $\alpha$; the same holds in section 7 for the time correlation functions with $r$ replaced by the time $t$ and $k$ by the frequency $\omega$. However, the singularities need not be branch points. 

In the present case of a vector field $\chi({\bf k})$ the singularity turns out to be a {\it discontinuity} at ${\bf k} = \{ k_x, k_y, \cdots \}=0$. This is shown by explicit calculations based on fluctuating hydrodynamics~\cite{Zia,DKS,Grinstein}
or ring kinetic theory~\cite{Buss+E}. The presence of the 
driving field makes the susceptibility anisotropic, i.e. 
\be
\lim_{{\bf k} \to \infty} \chi({\bf k}) = \chi_o(\hat{{\bf k}})
\ee
depends on the direction $\hat{{\bf k}}$ along which the origin is approached. 
The long range part of $G({\bf r})$ is then given by
\be
G({\bf r}) =  \frac{1}{(2 \pi)^d} \int d{\bf k} 
 \chi_o(\hat{{\bf k}}) \exp(i{\bf k\cdot r})
= \frac{h(\hat{{\bf r}})}{r^d} ,
\ee
where a rescaling of the ${\bf k}$--variable shows that the pair correlation
 function decays algebraically, with a coefficient $h(\hat{{\bf r}})$ 
that only depends on the direction of ${\bf r}$. An illustrative 
 landscape plot of $G({\bf r})$ can be found in Ref~\cite{Lebow}.
 Also, note the importance of the anisotropy. If $\chi_o$ were 
to be a constant, independent of $\hat{{\bf k}}$, then the correlation
function above would reduce to 
$G({\bf r}) = \chi_o \delta({\bf r})$, and is strictly short ranged. 
The algebraic tails $\sim 1/r^d$ are intimately related to the long time 
tails $\sim 1/t^{d/2}$, discussed above. In fact, they are driven by the 
same slow
diffusive modes of the ring kinetic equation, where spatial and temporal
scales are related by $r^2 \sim t$.

Also in the context of lattice gas automata the long range spatial 
correlations in velocity and density fields have been observed in simulations
over about 1.5 decades on the spatial scale, and the theoretical calculations
obtained from ring and repeated ring kinetic theory are in excellent 
agreement with the results from molecular dynamics 
simulations \cite{Buss+E}.

\section{Ring kinetic theory}
The "red thread" connecting all sections in this review are the ring 
collisions, which were born out of the systematic analysis
of the triple collision integrals of the Choh Uhlenbeck
equation. Hence the name ``cradle of modern kinetic 
theory''. Ring kinetic theory has been at the basis of most major 
developments
in nonequilibrium statistical mechanics over the last 40 years: it explains the
breakdown of the virial expansion of transport coefficients in continuous
fluids, and their logarithmic density dependence \cite{DvB}; it explains
the algebraic long time tails of Green Kubo current correlation functions
in continuous fluids \cite{Pom+R} and in lattice gas 
automata \cite{Brito}; the breakdown of the Navier Stokes equations in two
dimensions for very long times, as well as the non-analytic dispersion 
relations for sound propagation and for relaxation of hydrodynamic 
excitations~\cite{Pom+R}. Moreover, it explains the existence
of long range spatial correlations in nonequilibrium stationary states, coupled
to reservoirs that impose shear rates or temperature 
gradients~\cite{Onuki,DKS}, or in
driven diffusive systems~\cite{Grinstein,Zia}.

It seems therefore appropriate to give a simple derivation of the
ring kinetic equation for hard spheres, as given in Refs.~\cite{Ring-ED,Onuki}. Under
the restriction of low densities, the BBGKY hierarchy for hard spheres with
diameter $\sigma$ can be obtained from  (4) and (5) by replacing $\theta_{12}$ by the binary collision operator:
\be
T_0(12) = \delta (\vr_{12}) \sigma^{d-1} \int_{v_{12}\cdot\hat{\sigma} < 0}
d\hat{\sigma} |\vv_{12}\cdot\hat{\sigma}|(b_{\mbox{\boldmath $\sigma$}}-1) .
\ee
The vector $\mbox{\boldmath $\sigma$}$ connects the centers of the two
colliding spheres at contact and hats denote unit vectors. The integration over the
solid angle $d\hat{\sigma}$ is restricted to the precollision hemisphere,
$\vv_{12}\cdot\mbox{\boldmath$\sigma$} < 0$. The operator $b_{\mbox{\boldmath
$\sigma$}}$ replaces the precollision velocities $\vv_i$ $(i=1,2)$
by postcollision ones, i.e.
\bea
\vv^{*}_{1} & \equiv & b_{\mbox{\boldmath $\sigma$}} \vv_1 = \vv_1 +
(\vv_{12}\cdot\hat{{\mbox{\boldmath $\sigma$}}})\hat{{\mbox{\boldmath $\sigma$}}}
\nonumber \\
\vv^{*}_{2} & \equiv & b_{{\mbox{\boldmath $\sigma$}}} \vv_2 = \vv_2 -
(\vv_{12} \cdot\hat{{\mbox{\boldmath $\sigma$}}} 
)\hat{{\mbox{\boldmath $\sigma$}}}.
\eea
The Boltzmann equation for a dilute gas of hard spheres follows from the hard
sphere hierarchy by combining the first equation with the molecular chaos assumption
(2). As the density increases the ring collisions build up dynamic correlations,
which invalidate the molecular chaos assumption.

 The method is based on a cluster expansion of the
$s$-particle distribution functions, defined recursively as
\bea
f_{12} & = & f_1 f_2 + g_{12} \nonumber \\
f_{123} & = & f_1 f_2 f_3 + f_{1}g_{23} + f_{2}g_{13} + f_{3}g_{12} + g_{123} ,
\eea
and so on. Here $g_{12}$ accounts for pair correlations, $g_{123}$ for triplet 
correlations, etc. The molecular chaos assumption implies $g_{12} = 0$, which
is equivalent to (2). The basic assumption to obtain the ring kinetic
equation is that pair correlations are dominant and higher order terms in (27)
can be neglected, i.e. $g_{123} = g_{1234} = \ldots = 0$.

Setting $g_{12}=0$ in the first hard sphere hierarchy equation and
eliminating $\partial f_i / \partial t$ $(i =1,2)$ from the second
hierarchy equation with the help of the first one, yields the ring kinetic theory of elastic hard spheres:
\bea
 ( \partial_t &+& {\cal L}^0 (1)) f_1  = \int dx_2 T_0 (12) 
(f_1 f_3 + g_{12}) \nonumber \\
 \{ \partial_t + {\cal L}^0 (12)  &-&  T_0 (12) - (1+ {\cal P}_{12}) \int 
dx_3 T_0 (13) (1 + {\cal P}_{13})f_3 \} g_{12} \nonumber \\
 &=&   T_0 (12) \{f_1 f_2 + g_{12} \} .
\eea
Here ${\cal P}_{ij}$ is a permutation operator that interchanges the particle
labels $i$ and $j$. The second equation is the so called {\it repeated ring}
equation for the pair correlation function. If the operator $T_0(12)$ on
the left hand side of the second equation is deleted, one obtains the simple
{\it ring} approximation. Formally solving this equation for $g_{12}$ yields an
expression in terms of the single-particle distribution functions $f_i$
$(i=1,2,3)$, and subsequent substitution into the first hierarchy equation 
above yields the generalized Boltzmann equation in ring approximation.
Essentially all results about singularities and tails, discussed in sections 5 to 9, can be obtained from the 
ring kinetic equation, as given in (28). However, the {\it convergent}
Choh Uhlenbeck correction $n \eta_1$ to the three-dimensional viscosity in (15) is not correctly accounted for in this approximate equation.

\section{Molasses tails}
At liquid densities the velocity correlation function shows a 
negative part at intermediate times, 
the so-called {\it cage effect}~\cite{molass}. At the same density 
and in the same time interval the  stress correlation function in (7)
 appears to have in three dimensions a long time tail, that can be fitted to
$ t^{-3/2}$, but with a coefficient 400 to 500 times larger then
 predicted by the mode coupling theory of section 7. This so-called 
molasses tail occurs in dense liquids with short range hard core 
interactions, where the mean free path $\ell_0$ is much smaller 
than the hard core diameter $\sigma$.

The conventional {\it long} time tail $\sim t^{-d/2}$ is controlled by long 
wavelength modes, and the {\it intermediate} time molasses tail by
 slow modes of intermediate 
wavelength around $ k\sigma \simeq 2 \pi$, where the structure factor has 
its maximum. At these wavevectors the dynamics of fluid excitations 
 is in fact controlled by a single 
soft (density) mode, also called {\it cage diffusion}~\cite{Coh-Berl}. At the high densities relevant for liquids, the Fourier 
modes of momentum and energy density with wave vectors near $ 2 \pi / \sigma$ relax very rapidly through the mechanism 
of instantaneous collisional transfer over distances of order $\sigma$, 
whereas kinetic transport, which is operating on scales of 
 $\ell_o $, is much slower. Therefore the 
density, which can only relax through kinetic transport, is the only slow
 mode  near  $k \sigma \simeq 2 \pi$. It requires rearrangments of particle 
configurations over distances of order $\sigma$ (see review~\cite{Coh-Berl}),
{\it i.e.} structural relaxation. The detailed explanation was given by 
Kirkpatrick~\cite{Kirk-cage} and van Beijeren~\cite{HvB-cage},  who 
extended the mode coupling theory of section 7 by including 
wavenumbers upto $k \simeq 2\pi/ \sigma$, i.e. by including the slow cage
diffusion mode. This important fundamental result marks perhaps the end
of the modern era.

\section{Post-modern kinetic theory}
By 1985 the results of the modern era: the log $n$--dependence of transport 
coefficients, the long spatial and temporal tails in correlation functions, 
and the fact that  conventional hydrodynamics breaks down, at least in principle, were generally 
accepted. However, the experimental consequences of the exciting developments
 were minute and difficult to observe experimentally.

Where to go from here? The tendency was to combine intuitive classical 
concepts with the modern concepts of dynamical correlations, originating 
from mode coupling and ring kinetic theory, and to apply these to more 
complex problems, such as nonlinear hydrodynamics and porous 
media~\cite{6-gang}, dense fluids~\cite{Coh-Berl}, granular 
flows~\cite{granul}, mathematical modeling of the Boltzmann equation
and ring kinetic equations~\cite{DLB},
 and even to the calculation of Lyapunov exponents in 
non-equilibrium fluids~\cite{Lyap}. Simultaneously one sees the development of
 new methods for numerical simulation of non-equilibrium fluids, such as 
Lattice Gas Automata (LGA)~\cite{6-gang}, the Lattice Boltzmann Equation 
(LBE) method~\cite{LBE}, and  Bird-type methods~\cite{Bird}, which are 
stochastic simulation methods for solving nonlinear kinetic equations.

Many of these current developments, such as the LBE method, 
the Bird-type methods, mathematical modeling of kinetic equations, and 
Lyapunov exponents, 
have no direct links with the modern concepts of dynamical correlations 
and with the Choh Uhlenbeck equation, and will not be discussed any further.

An outburst of renewed interest in kinetic theory occurred in 1986, when
 Frisch, Hasslacher and Pomeau~\cite{6-gang} proposed Lattice Gas Automata (LGA),
 as models for non-equilibrium fluids. This event marks in my view the
 beginning of the {\it Post-modern Era}. The LGA, already defined in 
section 2, are fully discretized statistical mechanical models of 
$N$--particle systems, which do not involve any new fundamental concepts.
 The LGA were specially designed for large scale and long time computer 
simulations of complex hydrodynamic flow problems. As the dynamic 
description is based on Boolean variables, there are no round off errors.
The kinetic theory for these models closely parallels that for continuous 
fluids, as I have been indicating in all previous 
sections.

The most important fundamental result from LGA is the 
excellent confirmation of the theoretical predictions for the long time 
tails in the velocity correlation function, calculated from mode 
coupling  and ring kinetic theory. Here the
 tails have been measured with very high statistical accuracy over two 
decades of mean free times~\cite{Frenkel}.

Moreover, the study of LGA has provided many interesting new results on 
the dynamics of phase separation and pattern formation, as well as on 
flows through porous media (see review by Rothman and Zaleski~\cite{6-gang}). 
Unfortunately, LGA turned out to be not very suitable to study high Reynolds
 number flows, because the viscosity of LGA is too large.

A phenomenological equation linked to the modern era, and widely applied in
the post-modern era, is the {\it nonlinear} revised Enskog equation for hard spheres, which is a Markovian approximation,
 constructed in 1972 by van 
Beijeren and the author~\cite{RET-vBE} as an exact result for short times. 
The {\it 
linearized} version of this theory has been derived by many authors~\cite{RET}.
 These theories are collectively referred to 
as the {\it Revised Enskog Theory} (RET) with a collision operator taking
 non-local short range static correlations into account.

The theoretical and experimental research by de Schepper
and collaborators~\cite{Coh-Berl} on dynamical structure functions of dense liquids with 
short range hard core interactions has shown that the RET gives a 
reasonably accurate
description of the elementary excitations with wavelengths in the range 
$\ell_o < \lambda \leq \sigma$. A Fourier mode analysis~\cite{RET-modes} 
of the revised Enskog equation shows 
that this equation has only  a single slow mode at wavenumbers near $k \sigma 
\simeq 2 \pi$, the cage diffusion mode.
Although the RET involves only binary collision dynamics, it
 mimics the effects of cage diffusion, described in section 11, 
reasonably well. The classical Enskog equation in (3) is missing this effect.

Further interesting extensions of the RET to non-equilibrium solids 
have been given by Kirkpatrick et al~\cite{Kirk-solid}, who derived the 
macroscopic equations of motion for the elastic solid from this equation. 
RET yields in principle a microscopic theory for elastic constants and dissipative 
coefficients, as well as a dynamic theory for freezing. In the fluid phase
there are five slow hydrodynamic modes; in the solid phase there are eight 
soft modes 
with vanishing dispersion relations as $k \to 0$, i.e.  five conserved 
densities and three displacements fields (order parameter fields). There are
 six propagating modes, a thermal diffusion mode and a vacancy diffusion
 mode. Vacancy diffusion, which involves structural relaxation, occurs 
on time scales much greater than those occuring in sound phenomena and
 thermal diffusion, which are controlled by collisional transfer of
 momentum and energy. This suggests that cage diffusion in the  liquid 
phase and vacancy diffusion in the solid phase may be related to the same 
slow microscopic excitation.
Moreover, the RET has also been applied by  Kirkpatrick~\cite{Kirk-glass}, Kawasaki~\cite{Korea} and  
Dufty~\cite{Dufty-priv} to study the slow structural relaxation in glasses. 

The ring equation of section 10 has still a very complex structure. Dufty, 
Lee and Brey~\cite{DLB} have proposed a mathematical model with a  
simpler structure than (28). It would be very interesting to obtain 
from this simpler model more 
explicit analytic results for the pair correlation function in 
non-equilibrium fluids.
Finally, a very recent publication by van Noije et al.~\cite{PRL-us} on long range 
correlations in rapid granular flows shows that {\it ring kinetic theory} 
or fluctuating hydrodynamics has become the standard tool to analyze long 
range correlations in complex and simple fluids out of equilibrium. 

The central theme of this review has been the importance 
in non-equilib\-rium statistical mechanics of the ring 
collisions, which were born out of a detailed 
analysis of the Choh Uhlenbeck equation.

\section*{Acknowledments}
It is a pleasure to thank J.R. Dorfman, J.W. Dufty, R. Brito and T.P.C. van Noije 
for valuable comments, suggestions and help during the preparation of this
 manuscript. The author also thanks the University of Florida,
where part of this paper was written, for financial support.

\end{document}